\documentclass[journal,10pt]{IEEEtran}
\usepackage{amssymb}
\usepackage{amsmath}
\usepackage{cite}
\usepackage{url}
\usepackage{xcolor}
\usepackage{cite,graphicx,amsmath,amssymb}
\usepackage{subfigure}
\usepackage{citesort}
\usepackage{fancyhdr}
\usepackage{mdwmath}
\usepackage{mdwtab}
\usepackage{caption}
\usepackage{amsthm}
\usepackage{algorithm}
\usepackage{algorithmic}
\usepackage{graphicx} %use graph format
\usepackage{epstopdf}
\usepackage{setspace}
\usepackage{tikz}
\usepackage{amsmath, bm}
\usepackage{caption}
\usepackage{stfloats}
\usepackage{multirow}
\usepackage{tabularx}

\usepackage[justification=centering]{caption}

\newtheorem{remark}{Remark}
\newtheorem{theorem}{Theorem}
\newtheorem{definition}{Definition}

\newtheorem{lemma}{Lemma}

\newtheorem{corollary}{Corollary}

\makeatletter
\def\ScaleIfNeeded{%
\ifdim\Gin@nat@width>\linewidth \linewidth \else \Gin@nat@width
\fi } \makeatother

\begin{document}

%\title{\Huge{Mobile Edge Generation for Text-to-Image Generation Tasks}
%}

\title{\Huge{Enabling Distributed Generative Artificial Intelligence in 6G: Mobile Edge Generation}
}

\author{\normalsize {Ruikang~Zhong,~\IEEEmembership{\normalsize Member,~IEEE,}
Xidong~Mu,~\IEEEmembership{\normalsize Member,~IEEE,}\\
Mona~Jaber,~\IEEEmembership{\normalsize Senior~Member,~IEEE,}
Yuanwei~Liu,~\IEEEmembership{\normalsize Fellow,~IEEE.}\\
}

\thanks{
Ruikang~Zhong, Mona Jaber, and Yuanwei~Liu are with the School of Electronic Engineering and Computer Science, Queen Mary University of London, London E1 4NS, U.K. (e-mail: r.zhong@qmul.ac.uk; m.jaber@qmul.ac.uk; yuanwei.liu@qmul.ac.uk).
%Ruikang~Zhong and Mona Jaber are with the School of Electronic Engineering and Computer Science, Queen Mary University of London, London E1 4NS, U.K. (e-mail: r.zhong@qmul.ac.uk; m.jaber@qmul.ac.uk).
%
Xidong Mu is with the Centre for Wireless Innovation (CWI), Queen's University Belfast, Belfast, BT3 9DT, U.K. (e-mail: x.mu@qub.ac.uk)
%
%Yuanwei Liu is with the Department of Electrical and Electronic Engineering, The University of Hong Kong, Hong Kong (e-mail: yuanwei@hku.hk).
}
}

\maketitle
\begin{abstract}
Mobile edge generation (MEG) is an emerging technology that allows the network to meet the challenging traffic load expectations posed by the rise of generative artificial intelligence~(GAI). A novel MEG model is proposed for deploying GAI models on edge servers (ES) and user equipment~(UE) to jointly complete text-to-image generation tasks. In the generation task, the ES and UE will cooperatively generate the image according to the text prompt given by the user.  To enable the MEG, a pre-trained latent diffusion model (LDM) is invoked to generate the latent feature, and an edge-inferencing MEG protocol is employed for data transmission exchange between the ES and the UE.  A compression coding technique is proposed for compressing the latent features to produce seeds.  Based on the above seed-enabled MEG model, an image quality optimization problem with transmit power constraint is formulated. The transmitting power of the seed is dynamically optimized by a deep reinforcement learning agent over the fading channel. The proposed MEG enabled text-to-image generation system is evaluated in terms of image quality and transmission overhead. The numerical results indicate that, compared to the conventional centralized generation-and-downloading scheme, the symbol number of the transmission of MEG is materially reduced. In addition, the proposed compression coding approach can improve the quality of generated images under low signal-to-noise ratio (SNR) conditions.
\end{abstract}

%\begin{keywords}
%Deep learning, image generation, generative artificial intelligence, mobile edge generation.
%\end{keywords}

\section{Introduction}

Since the advent of GPT in 2022, the market and development of generative artificial intelligence have expanded exponentially. The generative artificial intelligence (GAI) service can customize the content according to the user's demand, and even complete complicated tasks on behalf of the users (e.g. CoPilot) \cite{9955525}. The number of users/applications using GAI models is increasing at a very high pace. After the invention of the telephone, it took about seventy years to acquire one million users. However, ChatGPT, a single GAI service provided by OpenAI, acquired one million users within only two months~\cite{10176168}. This emerging market has attracted the enthusiasm of IT companies. Microsoft, Google, and Alibaba have successively launched their GAI models \cite{saetra2023generative}. The data types supported by the GAI model are also expanding, including but not limited to text, audio, image, and videos~\cite{10256109}.

The emergence of GAI presents new challenges on the bandwidth and latency for communications networks \cite{10268594}. Not only does the GAI service have a large user base, but users also often do not receive a fully satisfactory generation product at the first attempt, resulting in repeated generation requests. Generally, currently available GAI services follow a request-generate-download model. To generate the product, the user first sends the generation request to the server where the GAI model is deployed, then the GAI model completes the generation task and downlinks the generated product to the user equipment (UE). Thus, this practice of high-frequency repeated requests leads to a considerable increase in network traffic, particularly when the produced content consists of high-definition images and videos \cite{aldausari2022video}. Taking ChatGPT as an example, as of February 2024, its page visits have accumulated to 1.7 billion. Therefore, the communication resources required by the frequent access to GAI services are substantial.  In addition, it is essential to emphasize that the quality of GAI service is determined by the network's transmission capacity to some extent. Transmission errors or delays can significantly impair the quality of the received content or increase waiting time for users. It follows that recently, a new research question related to communication systems has emerged to address the challenge imposed by GAI requirements for efficient and high-quality transmission between users and servers. Unfortunately, existing research on GAI often overlooks the transmission overhead caused by it, instead focusing on the overhead caused by the model training and calling.

\begin{figure*}[htb]
\centering
\includegraphics[width=0.85\textwidth]{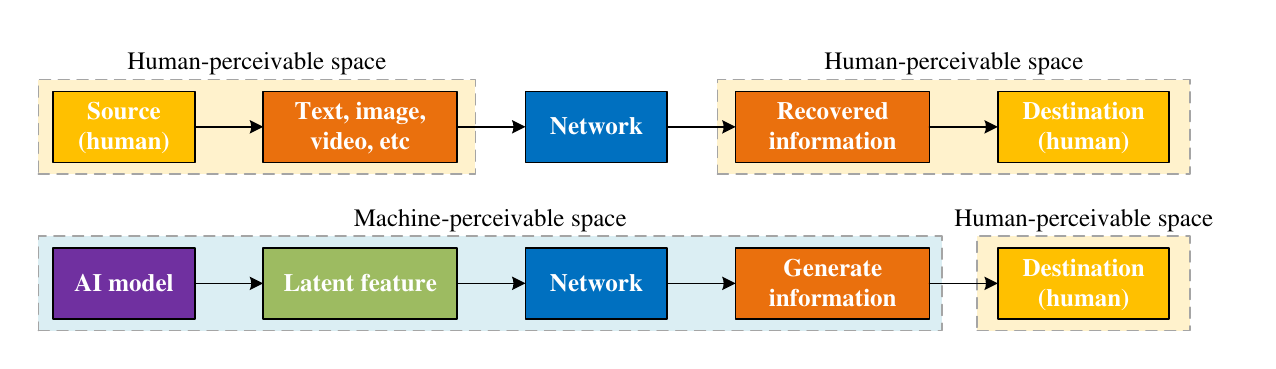}
\caption{The macroscopic impact of generative AI information sources on communication systems.}
\label{Fig.1}
\end{figure*}

We conducted research to provide an efficient communication solution for GAI, aiming to reduce communication overhead and ensure transmission quality. Inspired by mobile edge computing \cite{8764580}, we propose a mobile edge generation (MEG) solution to deploy the GAI model to edge devices in the network as a potential solution. The MEG closely integrates the GAI model with the edge network, deploying the GAI model to edge servers (ES) and UE. Information interaction between the GAI model deployed on ES and UE is essential for completing the generation task jointly. Therefore, how to carry out an effective and efficient communication scheme between ES and UE has become the scope of this paper. The proposed MEG framework focuses on solving the deployment and communication problems of GAI in the network. It can interoperate with a variety of GAI methods. However, this may require different specialized designs in terms of model deployment, encoding, etc.

\subsection{Research Motivation}

Although the network load brought about by the GAI service is significant, it is more crucial to note that the GAI model is likely to lead to a fundamental change in communications, rather than just an emerging application. The GAI model fundamentally changes the communication model by introducing a new type of information source. In any communication system model, the information source and information destination are the most basic components. However, it is worth noting that in conventional communication systems,  the final information source and destination are generally played by humans. Considering a phone call as an example, after the voice signal is generated by a human user, then it is processed by various technologies such as sampling, coding, and modulation. The signal also needs to be restored at the receiving end so that it can be heard by another user. Therefore, as shown in Fig. \ref{Fig.1}, the information has to be input and output in human-perceivable form at both ends of the network. Human-perceivable information is information that can be captured and analyzed by human senses, such as images, audio, etc. The information is in the encoded state only during the transmission, and it has to be operated in human-perceivable space at both the transmitter and receiver sides.

However, the convention was shattered with the advent of GAI, which became an intelligent source capable of producing original information. Observing Fig. \ref{Fig.1}, since the information source is replaced by the GAI model, it is no longer necessary to process the information in a human-perceivable space before it is finally presented to the user. This change becomes a prerequisite for redesigning communication protocols and coding techniques for GAI-engaged communication. The communication system that utilizes GAI as the source no longer needs to adhere to the process of encoding and restoring signals that are perceivable by humans. Specifically, in MEG, agents deployed on ES and UE can perform any necessary operations in the machine-perceivable space before finally being presented to the destination user. Machine-perceivable information refers to digital information that can be recognized and processed by computers and AI models. By removing the limitations of human perception, communication efficiency can be improved due to the new degrees of freedom. In this paper, we explore the use of encoded latent features as a means of communication between ES and UE, reducing the communication overhead of MEG and improving the GAI performance.

\subsection{The State-of-the-art Research}

\subsubsection{GAI for Image Generation}

There are several popular theories for the generation of images, including but not limited to the variational autoencoder (VAE) \cite{8638964}, \cite{8794603}, diffusion models \cite{10382547}. In 2019, a vector quantized variational autoencoder model was proposed in \cite{NEURIPS2019_5f8e2fa1} for the task of generating images having higher coherence and fidelity. By refining the features with fine-grained word embedding iteratively, a novel semantic-spatial aware GAN was proposed in \cite{Liao_2022_CVPR} for text-to-image synthesizing. The authors of \cite{Zhang_2022_CVPR} proposed to employ transformers to build a generative adversarial network for the generation of high-resolution images.

In recent years, diffusion models attracted the attention of researchers due to their outstanding generation abilities. A fundamental denoising diffusion probabilistic model (DDPM) was proposed in \cite{ho2020denoising}. After that, in order to reduce the complexity of the diffusion model, a denoising diffusion implicit model was proposed to accelerate the sampling process by constructing a class of non-Markovian diffusion processes~\cite{song2020denoising}. Based on the DDPM model, a denoising diffusion model for the multi-modality image fusion task was proposed in\cite{Zhao_2023_ICCV}. The diffusion models have a major limitation due to their complexity, especially for high-resolution images, since they require stepwise estimation and noise removal during the denoising process. To address this issue, the latent diffusion model (LDM) was proposed to reduce the complexity by operating the diffusion model in latent space \cite{Rombach_2022_CVPR,Wang_2023_CVPR}, which inspired the idea of operating transmission in the latent space in this paper.

\subsubsection{Communication and GAI Models}

The generative model was also invoked for resolving the communication problems in terms of signal detection \cite{sun2020generative}, and channel modeling~\cite{ye2022channel}. A generative adversarial network (GAN) approach was proposed in \cite{9169908} to compress and recover the feedback CSI to reduce the signaling overhead. In \cite{10328186}, generative models were designed to enable semantic communication, and the simulation results on data set CIFAR10 demonstrated that GAI enabled transmission can improve the Peak-Signal-to-Noise-Ratio (PSNR) for image transmission. Most recently, a concept of a generative network layer was proposed in \cite{10399967}, which employs GAI models at intermediate or edge network nodes to generate images from prompts instead of transmitting the image to improve communication efficiency.

\subsubsection{Distributed GAI}

Since the emergence of GAI, numerous recent surveys and magazine articles have introduced the possible combining of GAI and edge networks \cite{10422716,10398474}.
An edge artificial intelligence (Edge AI) model was proposed in~\cite{10384606}. The authors proposed a framework that can effectively coordinate edge AI models, leveraging large language models (LLM) to provide service for users. In order to resolve the issue caused by the resource-intensive nature of the GAI model, the authors of~\cite{10409284} introduced an AI-generated content service architecture, which deploys GAI models in the edge of wireless networks to enhance the accessibility of GAI services. The relation of the globe GAI model and the edge GAI model was also studied. In \cite{kusumaraju24ecogen}, the authors proposed a Generative AI-oriented synthetical network, which promotes a two-way knowledge flow, allowing pre-trained cloud GAI model to provide basic knowledge for edge models, while the edge models also aggregate personalized knowledge for GenAI. The privacy problem in the edge AI was considered in~\cite{9194756}. A NetGPT model was proposed that can collaborate with appropriate LLM on the edge and cloud-based on computing capabilities to provide personalized services to users~\cite{10466747}. In~\cite{10472660}, the authors proposed leveraging wireless sensing to guide GAI models to provide generation services in resource-limited mobile edge networks.

\subsection{Contributions}

The solution of invoking MEC to support GAI as referred to above has indeed made outstanding research contributions to solving the computational power problem of GAI. Some studies have also demonstrated the idea of deploying GAI models on edge devices. However, existing work does not fully consider the communication requirements of the GAI model and the impact of the introduction of the GAI model on the communication paradigm. Motivated by this knowledge gap, we propose the scheme of MEG to release the challenge for cellular networks caused by the emergence of the GAI.Based on the aforementioned analysis of how the GAI changes the protocol of communication, the research contributions of this work are summarized as follows:
\begin{itemize}
    \item We propose a novel MEG-enabled text-to-image generation model, which transmits the generation seed formed by coded latent features instead of the image itself. A latent diffusion model is invoked for generating the latent feature and a deep learning (DL) based coding technology is proposed for compressing the latent feature.
    \item A pair of DL-based compression encoder and decoder are designed to encode latent features into generation seed. This encoder can not only further compress the size of the latent feature that needs to be transmitted, but also enhance the generated image quality in low signal-to-noise ratio (SNR) conditions.
    \item Based on the proposed MEG framework and the coding technology, an image quality optimization problem is formulated. A deep reinforcement learning (DRL) approach is designed to dynamically allocate transmitting power for the seed transmission against the fading channel to optimize the quality of the generated image under a total power constraint.
    \item The simulation indicates that 1) the proposed MEG approach achieved higher image quality under low SNR conditions than other benchmarks; 2) employing generation seed significantly reduces the communication overhead compared to the centralized generation scheme; 3) the seed transmitted under the intelligent power control achieves significant image quality improvement compared to average power allocation.
\end{itemize}

%We first point out the problems, analyzing the wherefore of them, and finally modeling the effect of these phenomenons.

\subsection{Organization}

The rest of the paper is organized as follows: Section \ref{section:2} states the system model of MEG-enabled text-to-image generation. Section \ref{section:3} introduces the production of the generation seeds, where a pre-trained diffusion model is employed and a compression coding technology is proposed.  Based on the considered MEG seed transmission, Section \ref{section:4} formulates the image quality maximization problem, and a DRL-based power allocation algorithm is proposed for the formulated problem. Section \ref{section:5} demonstrates the numerical results to evaluate the image quality and the transmission overhead of the proposed MEG scheme compared to the conventional centralized generation scheme. Finally, the conclusion is provided in Section~\ref{section:6}.

\section{System Model}\label{section:2}

\begin{figure*}[t!]
\centering
\includegraphics[width=0.65\textwidth]{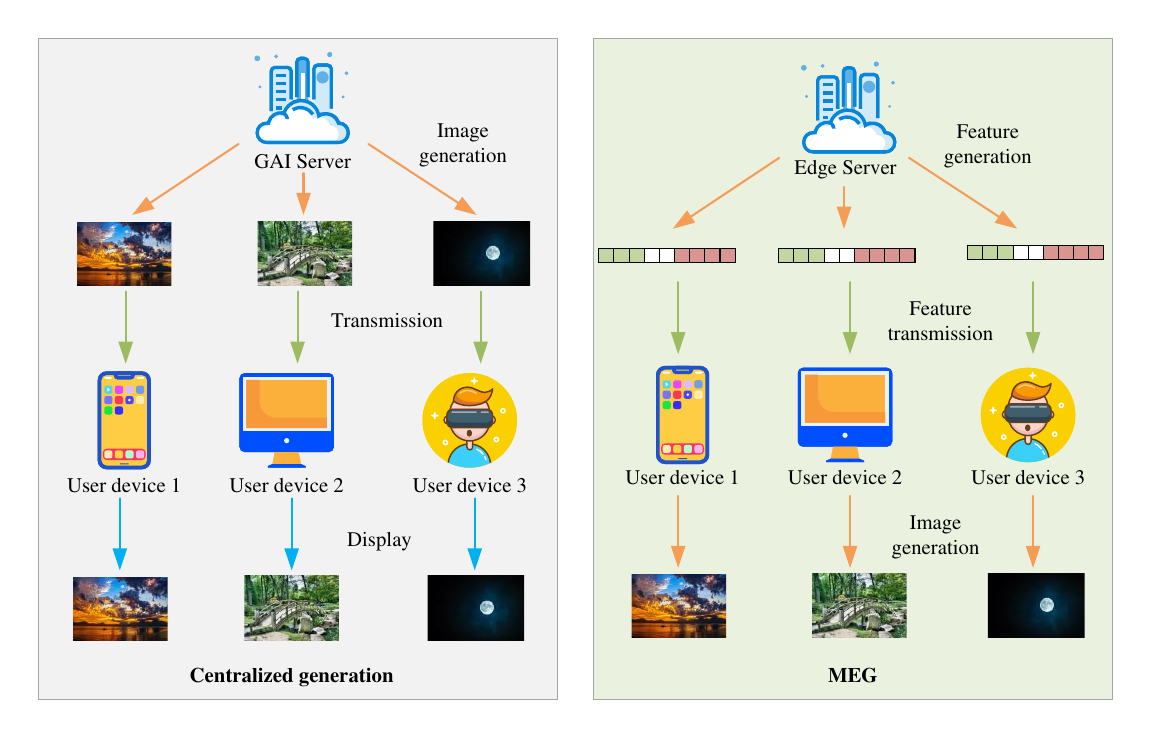}
\caption{System model of MEG.}
\label{Fig.2}
\end{figure*}

We consider a system consisting of an ES that provides text-to-image generation services, and UE requires the service by uploading the text prompt to the ES. ES is deployed at the base station of a cellular network and it is connected to GAI users through wireless channels. Fig. \ref{Fig.2} illustrated the model of the centralized generation and the MEG. In the case of conventional centralized generation, the GAI model is deployed on a server. When a user request is received, the image will be generated on the GAI server and then transmitted to the user through the wireless channel. On the contrary, in the case of MEG, an edge inferencing (EI) protocol~\cite{zhong2023mobile} is considered. The GAI model is partially deployed on ES and partially deployed on the UE, namely, the inferencing model and the generating model.  The inferencing model at ES has to infer the generation task, and the generating model at UE is responsible for the final image generation.

\subsection{Overview of MEG}\label{section:3A}

It is worth noting that MEG is not limited to any data modality of generated content nor the specific GAI algorithm deployed on the device, instead, it is a network technology that supports the completion of generation on network edge devices. The definition of MEG is provided in \textbf{Definition 1}.

\begin{definition}
Mobile edge generation is a network architecture concept that enables AI-based content generation capabilities at the edge of the cellular network and, more in general at the edge of any network.
\end{definition}
Specifically, the working flow of the EI protocol can be divided into three steps:
\begin{itemize}
\item \textbf{ES inference:} When the generation request is received, ES is responsible for generating a 'seed for the generation'\footnote{For the concise presentation, the term 'seed for the generation' are referred to as 'seeds'.} based on the prompt. This seed is an encoded feature that contains the information required for final generation.
\item \textbf{Seed download:} Once the seed is generated, it is transmitted to the UE by the ES through the wireless link. It is important to note that during transmission, seeds may be affected by noise and interference in non-ideal wireless channels.
\item \textbf{Final generation:} With the decoder and neural network model deployed on the UE, the UE will decode the seed and generate an image presenting to the user.
\end{itemize}

Due to the random noise and fading in the wireless channel, transmission errors are likely to be introduced during the download of images or seeds, thus affecting the quality of the image on the UE side. Therefore, we design a DL-based encoder and decoder for compression and error correction for the latent feature. Since a point-to-point transmission model is considered, we assume that there are no other interfering users. In addition, as the required bandwidth for transmitting text is minimal, we neglect the uplink transmission of the prompt and we assume that it can be received without any loss at ES.

\subsection{Signal Model}\label{section:3B}

Without loss of generality, we consider a user requesting a text-to-image generation task. The user for the generation service first enters the prompt string denoted by $\mathbf{s} = \{s_1, s_2 ...\}$ on the UE. After prompt $\mathbf{s}$ is sent to the ES, regardless of the specific generation algorithm, ES needs to perform two necessary operations from a systematic perspective. The first step is to use the inferencing model to generate features. Assuming that the inferencing model is denoted as $\mathcal{G}$, this process can be expressed as follows:
\begin{align}
 {{\bf z}} =  \mathcal{G}_{\bm{\alpha}}(\mathbf{s}, {\bf z}_0, f_d),
\end{align}
where $\mathbf{z}$ represents the generated feature, and $\bm{\alpha}$ represents the parameters employed in the model $\mathcal{G}$. ${\bf z}_0$ represents a feature formed by random noise. According to the definition in \cite{Rombach_2022_CVPR}, the vector ${\bf z}$ is defined as the latent feature, which refer to the underlying characteristics that can be extracted from the original image but cannot be directly observed. $f_d$ is the down sampling factor from the pixel space to the latent space. In the proposed MEG scheme, the information being transmitted is carried by the latent features. In this specific scenario, the latent feature is belongs to latent space (machine-perceivable) and the image is in pixel space (human-perceivable). The MEG aims to process the generation in the latent space and transfer it to the pixel space on user equipment.

After generating the feature $\mathbf{z}$, it is a feasible solution to directly downlink the generated feature to the UE. However, in order to pursue further compression and noise resistance, features need to be further encoded. Therefore, a compression coding scheme based on DL is proposed. The compression encoder is deployed in the ES and the decoder is deployed in the UE. The coded seed $\mathbf{x}$ can be given by
\begin{align}
 {{\bf x}} =  \mathcal{C}_{\bm{\omega}}(\mathcal{G_{\bm{\alpha}}}(\mathbf{s}), f_c),
\end{align}
where $\mathcal{C}$ and $\bm{\omega}$ represent the DNN model and its parameters for the compression encoding, respectively. $f_c$ is the compression rate for the feature. For different compression rate settings, the compression encoder and decoder also need to be trained accordingly. After the downlink transmission, the seed can be received by the UE, and the received signal can be given according to the classic point-to-point transmission model, which is
\begin{align}\label{y}
 {{\bf y}} = h{{\bf x}} + {{\bf n}},
\end{align}
where $h$ is the channel fading and ${\bf n}$ represents the additive white Gaussian noise (AWGN). At the receiver side, an opposite operation to that at ES needs to be performed. First use the decoder to recover the feature, which can be expressed as
\begin{align}
 {{\bf \hat{s}}} = \mathcal{C}^{-1}_{\bm{\psi}}(\mathbf{\hat{x}}, f_c),
\end{align}
where $\mathcal{C}^{-1}$ represents the decoder and $\bm{\psi}$ represents the parameters. Finally, the image can be generated according to the feature, which can be given by
\begin{align}
{{\bf I}_g} = \mathcal{D}_{\bm{\beta}}(\mathcal{C}^{-1}_{\bm{\psi}}(\mathbf{s}, f_c), f_d),
\end{align}
where ${{\bf I}_g}$ is the generated image and $\mathcal{D}$ represents the generation model which is responsible to generate image from the feature.

\begin{remark}
Transmitting in latent space requires less symbol cost. The dimensions of the latent feature and the image follow a relationship of $|\mathbf{z}| < f_d^{-2}|{{\bf I}_g}|, f_d >1$. Therefore, we have the relationship of transmitted symbol number $|\mathbf{x}| < |\mathbf{z}| << |{{\bf I}_g}|$, which ensures that using the MEG operating in a latent space can reduce the communication overhead.
\end{remark}

\subsection{Performance Metrics}\label{section:3D}

In this work, we evaluate the quality of MEG products by comparing the images generated by MEG with images generated under perfect conditions. We employ the peak signal-to-noise ratio (PSNR) \cite{winkler2008evolution} and Frechet Inception Distance (FID) score to evaluate the quality of the generated images.  We assume that there is an image generated under perfect conditions and we denote the perfect image as $\mathbf{I}^0$ as a ground truth. In the perfect generation condition, we assume there are zero transmission errors and sufficient computing power. We measure the quality of the generated image by comparing the generated images $\mathbf{I}^g$ with the perfectly generated image $\mathbf{I}^0$.

\subsubsection{PSNR}

PSNR was introduced to measure the difference between perfectly generated images and actual generated images. First calculate the mean square error (MSE) between $\mathbf{I}^0$ and $\mathbf{I}^g$ as
\begin{equation}
\text{MSE}_{\mathbf{I}^g} = \frac{\sum\limits_{m, n}^{M,N} {{{({\mathbf{I}^g} - {\mathbf{I}^0})}^2}}}{M*N},
\end{equation}
where $M$ and $N$ are the number of rows and columns in the input images.

\begin{equation}
\text{PSNR}_{\mathbf{I}^g} = 10\log 10\left( {\frac{{i_\text{max}^2}}{{{\text{MSE}_{\mathbf{I}^g}}}}} \right),
\end{equation}
where $i_\text{max}$ represent the maximum fluctuation in the image, which is also known as the highest possible pixel value (e.g. for the image using 8-bit unsigned integer data type, $i_\text{max}$ is 255).

\subsubsection{FID score}

Since PSNR can only determine the pixel difference between the actual generated image and the perfectly generated image, in order to better evaluate the generation performance, we introduce the FID score. The FID score can effectively calculate the distance between perfectly generated images and actual generated samples. The FID score was proposed and used by Martin Heusel \textit{et. al.} \cite{heusel2017gans}. The FID score is calculated by first loading the pre-trained Inception v3 model. The Inception v3 model takes an image as input and predicts an activation feature map for that image. Each image is predicted as an activation feature with a length of 2,048. The FID score is then calculated by
\begin{equation}
\text{FID}_{\mathbf{I}^g}  = \lVert \mu _{\mathbf{I}^g} - \mu _{\mathbf{I}^0}\rVert ^{\mathbf{I}^g} + \text {Tr}(C_{\mathbf{I}^g} + C_{\mathbf{I}^0} - 2\sqrt {C_{\mathbf{I}^g} \cdot C_{\mathbf{I}^0}}),
\end{equation}
where $\mu _{I^g}$ and $\mu _{I^p}$ represent the means of the feature vectors of perfect and generated images, and $C_{I^g}$, $C_{I^p}$ represent the covariance matrix for feature vectors of the perfect and generated images, respectively. Therefore, a smaller FID score indicates that the generated image is closer to the ideal image.

\begin{figure*}[t!]
\centering
\includegraphics[width=0.8\textwidth]{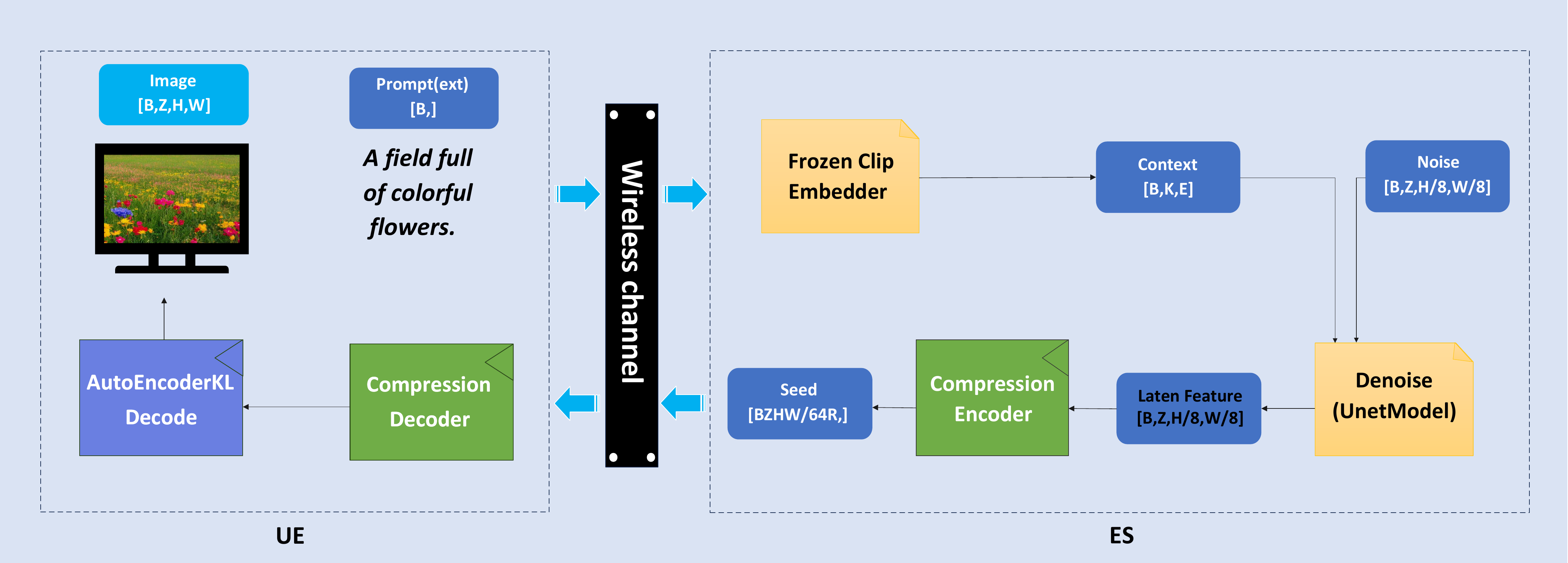}
\caption{MEG model of LDM-enabled text-to-image generation.}
\label{Fig.3}
\end{figure*}

\section{Seed Generation: Image MEG based on Stable Diffusion}\label{section:3}

We built a MEG generation scheme based on the Latent Diffusion Models proposed in \cite{Rombach_2022_CVPR}. LDM was selected as the generation scheme since it performs the denoising operation in the latent space.     The original motivation for using latent space in LDM was to simplify the denoising process. This architecture aligns perfectly with MEG's objective of letting as much operation as possible be executed in the machine-perceivable space. Fig.~\ref{Fig.3} reveals the structure and main components of the proposed MEG approach. The proposed solution mainly contains five parts, including a prompt embedder, a denoising model, a pair of compression encoder/decoder, and an autoencoder. The prompt embedder, denoising model, and autoencoder are pre-trained models, and we will focus on the training of the compression encoder and decoder.

\subsection{Prompt Embedding and Auto Encoder}

The text prompt  $\mathbf{s}$ uploaded to ES first need to be processed by CLIP Embedder into embedded vector, which can be described as
\begin{align}
{{\bf \tilde{s}}} =  \mathcal{B}_\mathbf{b}(\mathbf{s}), {{\bf \tilde{s}}} \in \mathbb{C}^{B\times K \times E}, \label{eq:embedding}
\end{align}
where $B$ represents the batch size, $K$ represents the maximum encoding length of text, and $E$ represents the size of embedding. Without loss of generality, a CLIP embedder $\mathcal{B}_\mathbf{b}(\cdot)$ is employed in this work.

Auto encoder is a key technology in LDM, which allows arbitrary images to be mapped from pixel space to latent space. As suggested in \cite{Rombach_2022_CVPR}, we employ a auto encoder to map the image $\mathbf{I}$ in to $\mathbf{z} = \mathcal{E}(\mathbf{I}), \mathbf{I} \in \mathbb{C}^{B \times C \times H \times W},  \mathbf{z}\in \mathbb{C}^{B \times Z \times H/f_d \times W/f_d}$, where $Z$ represents the channel number of the latent space.  A KL regularization is used in the auto encoder in order to avoid arbitrarily high-variance latent spaces, which imposes a slight KL-penalty a standard normal on the learned latent. The training goal of the auto encoding is given by equation \eqref{lossautoencoder} according to \cite{kingma2013auto}.
\begin{figure*}[!b]
\hrulefill
\centering
\begin{align}\label{lossautoencoder}
L _ \text{ Autoencoder } = \lim _ { \mathcal{E} , \mathcal{D} } \max _ { \psi } \left( L _ \text{rec} ( \mathbf{I} , \mathcal{D} ( \mathcal{E} ( \mathbf{I} ) ) ) - L _ \text{adv} ( \mathcal{D} ( \mathcal{E} ( \mathbf{I} ) ) ) + \log \mathcal{D} _ { \psi } ( \mathbf{I} ) + L _ \text{reg} ( \mathbf{I} ; \mathcal{E} , \mathcal{D} ) \right).
\end{align}
\end{figure*}

The term $L _ \text{rec}$ represents the loss to minimize the reconstruction error; $L _ \text{adv}$ represents the adversarial loss \cite{Esser_2021_CVPR}, and $L _ \text{reg}$ is the regularizing loss term to enforce the latent feature $\mathbf{z}$ to have zero center.
In the proposed MEG scheme, the autoencoder is deployed at the ES as shown in Fig. \ref{Fig.3}. The auto decoder $\mathcal{D}(\cdot)$ is deployed at the UE for the final image generation. While the UE recovered the received latent feature $\mathbf{\hat{z}}$, the auto decoder is responsible for generating the image by $\mathbf{I}_g = \mathcal{D}(\mathbf{\hat{z}})$.

\begin{figure*}[!b]
\centering
\begin{align}\label{qnonMarkov}
 q  ( \mathbf{z} _ { t - 1 } | \mathbf{z} _ { t } , \mathbf{z} _ { 0 } ) = N ( \sqrt { \alpha _ { t - 1 } } \mathbf{z} _ { 0 } + \sqrt { 1 - \alpha _ { t - 1 } - \sigma _ { t } ^ { 2 } } \cdot \frac { \mathbf{z} _ { t } - \sqrt { \alpha _ { t } } \mathbf{z} _ { 0 } } { \sqrt { 1 - \alpha _ { t } } } , \sigma _ { t } ^ { 2 } \mathbb{I} ) .
\end{align}
\end{figure*}

\subsection{Diffusion Model}

Diffusion models are a sort of generative model that aims to generate data that is similar to the training data. The diffusion model contains a diffusion process and a denoising process.  In the diffusion process, we sample a data point
$\mathbf{z}_0$ from the true data distribution $q(\mathbf{z})$ and then generate add isotropic Gaussian noise with a variance schedule $[\beta_1, ... \beta_T ]$, where $T$ represents the time step \cite{song2020denoising}. The set of sample $\mathbf{z}_1, \mathbf{z}_2 ... \mathbf{z}_T$ can be obtained using a predefined noise schedule.
\begin{align}
q(\mathbf{z}_{1: T} \vert \mathbf{z}_0):=q(\mathbf{z}_T|\mathbf{z}_0)\prod_{t=2}^T q(\mathbf{z}_t \vert\mathbf{z}_{t-1}).
\end{align}

This forward process which also known as diffusion process of gradually adding Gaussian noise to the data can be formulated into a Markov chain, which is given by
\begin{align}
q(\mathbf{z}_t \vert \mathbf{z}_{t-1})=\mathcal{N}(\mathbf{z}_t; \sqrt{1-\beta_t} \mathbf{z}_{t-1}, \beta_t \mathbb{I}),
\end{align}
where $\mathbb{I}$ represents identity matrix. In the DDIM sampler, the diffusion forward process is reformulated into a non-Markov format, which is given by
\begin{align}
q ( \mathbf{z} _ { T } | \mathbf{z} _ { 0 } ) = \mathcal{N} ( \sqrt { \alpha _ { T } } \mathbf{z} _ { 0 } , ( 1 - \alpha _ { T } ) \mathbb{I} ),
\end{align}
where $\alpha _ { T } = 1-\beta_T $. Then, for all $t>1$, the non-Markov forward process can be rewriter as equation \eqref{qnonMarkov}.

The opposite process of diffusion is denoising. During the denoising process, the noise in $q(\mathbf{z}_T)$ needs to be gradually eliminated to obtain latent features similar to $q(\mathbf{z}_0)$. Thus, a neuronal network model needs to be trained to estimate the reserve  $p(\mathbf{z}_{t-1}\vert \mathbf{z}_{t})$ of $q(\mathbf{z}_{t}\vert \mathbf{z}_{t-1})$. A generative process using trainable neuronal network $\theta$ can be denote as $ p _ { \theta } ( \mathbf{z} _ { 0 } : T )$. In each denoising step $t$, $ p _ { \theta } ( \mathbf{z} _ { t - 1 } | \mathbf{z} _ { t } )$  learn form the process of $q ( \mathbf{z} _ { t - 1 } | \mathbf{z} _ { t } , \mathbf{z} _ { 0 } ).$

For a given noise latent feature $ \mathbf{z} _ { 0 } \sim q ( \mathbf{z} _ { 0 } )$ and Gaussian noise $\epsilon _ { t } \sim N ( 0, \mathbb{I} ),$ the denoised observation, which is a prediction of noise-free feature $\mathbf{z}_0$ with given $\mathbf{z}_t$ can be calculated as
\begin{align}
f _ { \theta,t }  ( \mathbf{z} _ { t } ) : = ( \mathbf{z} _ { t } - \sqrt { 1 - \alpha _ { t } } \cdot \epsilon _ { \theta,t }  ( \mathbf{z} _ { t } ) ) / \sqrt { \alpha _ { t } } .
\end{align}

The process of a denoising step can be given by
\begin{equation}
 p_{\theta}(\mathbf{z}_{t-1}\vert \mathbf{z}_{t})=\mathcal{N}(\mathbf{z}_{t-1};f_{\theta}(\mathbf{z}_{t},\ t),\ \Sigma_{\theta}(\mathbf{z}_{t},\ t)).
\end{equation}

The performance of the generation is depended on the accuracy of the estimation of $\epsilon$. The U-Net is trained to estimate the $\epsilon$ by optimizing the Evidence Lower Bound (ELBO), where the ELBO over the discrete denoising steps is given by \eqref{ELBO}.
\begin{figure*}[!b]
\hrulefill
\begin{align}
- \log p ( \mathbf{z} _ { 0 } ) \leq K L ( q ( \mathbf{z} _ { T } | \mathbf{z} _ { 0 } ) | p ( \mathbf{z} _ { T } ) ) + \sum _ { t = 1 } ^ { T } E _ { q ( \mathbf{z} _ { t } | \mathbf{z} _ { 0 } ) } K L ( q ( \mathbf{z} _ { t - 1 } | \mathbf{z} _ { t } , \mathbf{z} _ { 0 } ) | p ( \mathbf{z} _ { t - 1 } | \mathbf{z} _ { t } ) ). \label{ELBO}
\end{align}
\end{figure*}

As $p_\theta ( \mathbf{z} _ { T } )$ is generally a standard normal distribution, in order to minimize the other terms, it is beneficial to target at a specific estimated $ \mathbf{z} _ { \theta } ( \mathbf{z} _ { t } , t )$ instead of a random $ \mathbf{z} _ { 0 } $. Then, the denoising step $p_\theta ( \mathbf{z} _ { t - 1 } | \mathbf{z} _ { t } )$ can be given by
\begin{align}
p_\theta ( \mathbf{z} _ { t - 1 } | \mathbf{z} _ { t } ) : = q ( \mathbf{z} _ { t - 1 } | \mathbf{z} _ { t } , \mathbf{z} _ { 0 } ( \mathbf{z} _ { t } , t ) ), \\ = \mathcal{N} ( \mathbf{z} _ { t - 1 } | \mu _ { t } ( \mathbf{z} _ { t } , t ) , \sigma _ { t + - 1 } ^ { 2 } \frac { \sigma _ { t - 1 } ^ { 2 } } { \sigma _ { t } ^ { 2 } } ] ) ,
\end{align}
and the mean of the noise is given by
\begin{align} \mu _ { \theta } ( \mathbf{z} _ { t } , t ) = \frac { \alpha _ { t | t - 1 } \sigma _ { t - 1 } ^ { 2 } } { \sigma _ { t } ^ { 2 } } \mathbf{z} _ { t } + \frac { \alpha _ { t - 1 } \sigma _ { t | t - 1 } ^ { 2 } } { \sigma _ { t } ^ { 2 } } \mathbf{z} _ { \theta } ( \mathbf{z} _ { t } , t ) .
\end{align}

Then, the ELBO can be rewritten as in \eqref{ELBO2}.
\begin{figure*}[!b]
\begin{align}
\sum _ { t = 1 } ^ { T } E _ { q ( \mathbf{z} _ { t } | \mathbf{z} _ { 0 } ) } ] K L ( q ( \mathbf{z} _ { t - 1 } | \mathbf{z} _ { t } , \mathbf{z} _ { 0 } ) | p ( \mathbf{z} _ { t - 1 } ) = \sum _ { t = 1 } ^ { T } E _ { N ( \epsilon | 0 , 1 ) } \frac { 1 } { 2 } ( \frac { \alpha _ { t-1 } ^ { 2 } } { \sigma _ { t-1 } ^ { 2 } } - \frac { \alpha _ { t } ^ { 2 } } { \sigma _ { t } ^ { 2 } } ) | | \mathbf{z} _ { 0 } - \mathbf{z} _ { \theta } ( \alpha _ { t } \mathbf{z} _ { 0 } + \sigma _ { t } \epsilon , t ) | | ^ { 2 }. \label{ELBO2}
\end{align}
\end{figure*}
According to~\cite{ho2020denoising}, by the reparameterization of
\begin{align}
\epsilon _ { \theta } ( \mathbf{z} _ { t } , t ) = ( \mathbf{z} _ { t } - \alpha _ { t } \mathbf{z} _ { \theta } ( \mathbf{z} _ { t } , t ) ) / \sigma _ { t },
\end{align}
the denoising objective is given by
\begin{align}  | | \mathbf{z} _ { 0 } - \mathbf{z} _ { \theta } ( \alpha _ { t } \mathbf{z} _ { 0 } + \sigma _ { t } \epsilon , t ) | | ^ { 2 } = \frac { \sigma _ { t } ^ { 2 } } { \alpha _ { t } ^ { 2 } } | | \epsilon - \epsilon _ { \theta } ( \alpha _ { t } \mathbf{z} _ { 0 } + \sigma _ { t } \epsilon , t ) | | ^ { 2 }.
\end{align}

Finally, the object function of training can be given by
\begin{equation}
{L^{{\text{LDM}}}} = {\mathbb{E}_{t,\mathbf{z},{{\bf \tilde{p}}},\epsilon }}{[\left\| {\epsilon - {{\mathbf{\epsilon }}_{{\theta}} }\left( {{\mathbf{z}_t},{{\bf \hat{s}}},t} \right)} \right\|^2_2]}.
\end{equation}
where $\epsilon$ is a random Gaussian noise added to $\mathbf{z}$, and $\epsilon_{\bm{\theta}}$ denotes the diffusion model trained to estimate the injected noise using optimized parameters ${{\theta}}$. In this work, we employ a pre-trained Unet model for the denoising process, and the model can be denoted as $\mathcal{U}_{{\theta}}$.

In the task of inference, the latent feature $\mathbf{z}$ is generated by iteratively operating a diffusion process $\mathbf{z}_t \rightarrow \mathbf{z}_{t-1}$ of a well-trained DDIM sampler by \eqref{sampling}. This sampling process requires high computing power and graphics processing unit (GPU) memory, thus, this process is assigned to ES processing because ES generally has stronger hardware facilities.
\begin{figure*}[!b]

\begin{align}
\mathbf{z} _ { t - 1 } = \sqrt { \alpha _ { t - 1 } } ( \frac { \mathbf{z} _ { t } - \sqrt { 1 - \alpha _ { t } } \epsilon _ { \theta } ^ { ( t ) } ( \mathbf{z} _ { t } ) } { \sqrt { \alpha _ { t } } } ) + \sqrt { 1 - \alpha _ { t - 1 } - \sigma _ { t } ^ { 2 } } \cdot \epsilon _ { \theta } ^ { ( t ) } ( \mathbf{z} _ { t } ) + \sigma _ { t } \epsilon _ { t }.\label{sampling}
\end{align}
\end{figure*}

\subsection{Compression Encoding and Decoding}
Compression encoding and decoding have two main functions. The primary function of Compression encoding is to further compress the size of latent features, thereby further reducing communication overhead. Secondly, with the help of compression encoding and decoding, the noise resistance performance of latent features is improved.
\subsubsection{Compression Encoder}

The compression encoder takes the latent feature from the LDM as the input. The compression encoder consists of a flatten layer followed by dense layers. The output code length and the compression rate $f_c$ can be adjusted by changing the number of nodes at the output layer. For different compression rates, the encoder may have different optimal structures and sizes. An example structure of the proposed compression encoder is given in \textbf{Appendix A}.

%\begin{figure}[t!]
%\centering
%\includegraphics[width=0.25\textwidth]{Figs/Encoder.pdf}
%\caption{Compression Encoder for feature transmission.}
%\label{Fig.4}
%\end{figure}
%
%\begin{figure}[t!]
%\centering
%\includegraphics[width=0.3\textwidth]{Figs/Decoder.pdf}
%\caption{Compression Decoder for feature transmission.}
%\label{Fig.5}
%\end{figure}

\subsubsection{Compression Decoder}

The compression decoder response for decoding the seed into the latent feature as $\hat{\mathbf{z}} = \mathcal{C}^{-1}_{\bm{\psi}}(\hat{\mathbf{x}})$. It consists of multiple dense layers, normalization layers, a residual layer, and an unfaltten layer. We use conventional RELU as the activation function. The empirically optimal decoder structure has three dense layers with normalization layers inserted in between. The full process of the MEG image generation is summarized in  \textbf{Algorithm 1}.
\begin{algorithm}
\caption{MEG Generation}
\label{train}
\begin{algorithmic}[1]
        \STATE \textbf{Initialization:}Initialize the trained models $\mathcal{B}_\mathbf{b}(\cdot), \mathcal{E}(\cdot)$, $ \mathcal{D}(\cdot)$ , $\mathcal{U}_{\bm{\theta}}, \mathcal{C}_{\bm{\omega}}(\cdot),$ and $ \mathcal{C}^{-1}_{\bm{\psi}}(\cdot)$.
        \FOR{Each generation task}
        \STATE Uploading the prompt $\mathbf{s}$.
        \STATE Embedding the prompt $\mathbf{s} \rightarrow \mathbf{\tilde{s}}$ by \eqref{eq:embedding}.
        \STATE Unet $\mathcal{U}_{\bm{\theta}}$ generate the feature according to prompt $\mathbf{\tilde{s}} \rightarrow \mathbf{z}$.
        \STATE Compression encoder $\mathcal{C}_{\bm{\omega}}(\cdot)$ perform $\mathbf{z}\rightarrow \mathbf{x}$.
        \STATE Seed download $\mathbf{x}\rightarrow \mathbf{y}$.
        \STATE Compression decoder $\mathcal{C}^{-1}_{\bm{\psi}}(\cdot)$ perform $\mathbf{\hat{x}}\rightarrow \mathbf{\hat{z}}$.
        \STATE Auto decoder $\mathcal{D}_{\bm{\beta}}(\cdot)$ recover the image form latent feature $\mathbf{\hat{z}}\rightarrow \mathbf{I}_g$.
        \ENDFOR
        \STATE \textbf{Output:} Image $\mathbf{I}_g$.
\end{algorithmic}
\end{algorithm}

\subsubsection{Joint Training}

In the task of image generation using MEG, the overall objective is to optimize the quality of the generated images by training the pair of compression encoder $\mathcal{C}_{\bm{\omega}}$ and decoder $\mathcal{C}_{\bm{\psi}}^{-1}$. Therefore, the optimization goal for the compression encoder and decoder is to minimize the transmission error caused by the noise and fading channels. For a given generation task, the problem of the compression coding can be formulated as
\begin{subequations}
\begin{align}
\mathbf{P1}: &\min_{\bm{\omega}, \bm{\psi}}   {\mathbb{E}_{\mathbf{z},\mathbf{\hat{z}}} }{[\left\|\mathbf{z} - \mathbf{\hat{z}} \right\|^2_2]}, \\
\textrm{s.t.} \ \
& 0 < |\mathbf{x}_{\mathbf{I}^g}| < f_d^{-2}|\mathbf{x}_{\mathbf{I}^c}|, f_d > 1  ,\label{OPP4}
\end{align}
\end{subequations}
where constraint (\ref{OPP4}) is a constraint on the number of the transmitted symbols. $|\mathbf{x}_{\mathbf{I}^g}|$ and $|\mathbf{x}_{\mathbf{I}^c}|$ represent number of symbols needed for the transmission in MEG approach and centralized generation approach, and $f_d$ is the down sampling factor from the pixel space to the latent space. This constraint can guarantee that the compression rate of the compression encoder is in the range of $(0,1)$.

We adopt the training of the compression encoder independently of the GAI model, which aims to recover the latent feature at the UE. Although the encoder and decoder pairs are deployed on different devices, they need to be trained jointly. First, the training data needs to be prepared for use with the LDM. The GAI model needs to generate a latent feature data set and save it. We use a perfect channel to train the encoder and the decoder, aiming to minimize the MSE of the original latent feature and the recovered feature. The loss function of the compressing coding can be given by
\begin{equation}
{L^{{\mathcal{C},\mathcal{D}}}} = {\mathbb{E}_{\mathbf{z},\mathbf{\hat{z}}} }{[\left\|\mathbf{z} - \mathbf{\hat{z}} \right\|^2_2]}, \label{eq:loss}
\end{equation}
where $\mathbf{\hat{z}}$ represents the decoded latent feature. The whole training process of the encoder $\mathcal{C}$ and decoder $\mathcal{D}$ is given in \textbf{Algorithm 2}.
\begin{algorithm}
\caption{Training Algorithm for the Compression Coding}
\label{train}
\begin{algorithmic}[1]
        \STATE \textbf{Initialization:}Generate the latent feature data set $\mathbf{Z} = {\mathbf{z}_1, \mathbf{z}_2 ...}$
        \STATE \textbf{Initialization:}Initialize the network $\bm{\omega}_{0}$, $\bm{\psi}_{0}$.
        \FOR{Each Epoch}
        \STATE Download the seed over fading channel with normalized power, obtaining $\mathbf{z},\mathbf{\hat{z}}$.
        \STATE Joint train the network $\bm{\omega}$ and $\bm{\psi}$ according to \eqref{eq:loss}.
        \ENDFOR
        \STATE \textbf{Output:} $\mathcal{C}_{\bm{\omega}}(\cdot)$ and $\mathcal{C}^{-1}_{\bm{\psi}}(\cdot)$.
\end{algorithmic}
\end{algorithm}

It is an irrefutable fact that the advent of compression encoders and decoders has resulted in an increase in computational complexity overhead. Nevertheless, the source encoding of centrally generated images also necessitates a considerable amount of computation. As the compression encoder and decoder are only required to process latent features, rather than the entire image, the computational complexity of MEG is not inferior to the centralized generation.

\subsection{Discussion}

A major advantage of the complexity of the proposed scheme comes from keeping learning and communication performed in the latent space until the image needs to be displayed on the UE. From a data transmission cost perspective, if the traditional generation-and-transmission scheme is employed, the amount of unencoded original data for in image $\mathbf{I}$ can be denoted as $|\mathbf{I}|$. However, for the MEG scheme, the amount of the transmitted data needs to be compressed twice. First, the image is transferred from pixel space to latent space, and it is compressed with a compression rate of $f_d$ in both height and width dimensions, and thus we have $|\mathbf{z}| = f_d^2 |\mathbf{I}|$.  After compression encoding, the length of the transmitted vector is $|\mathbf{x}| = f_c^{-1} f_d^{-2} |\mathbf{I}|$, which is a transmission complexity gain brought by applying MEG. It is true that other compression coding techniques (e.g. joint photographic experts group code) or error correction coding techniques (e.g. Turbo code, low-density parity check) can be used on the original image. It is worth emphasizing that these techniques can also be used for seed transmission. For the MEG system, encoding is a critical component that determines performance, and the MEG system allows a variety of encoding technologies to be applied once beneficial.

In terms of DL-enabled compression encoding, in this paper we employed a pre-trained GAI model and we trained the compression encoder and decoder pair separately from the GAI model. We have not fully explored joint training of encoders with GAI models, and we believe this may further improve the quality of the generated images.

The seed transmission-enabled MEG can not only be employed in point-to-point transmission, but it also has the potential to be extended to multi-user or multi-cell scenarios. In accordance with the proposed seed transmission scheme, a simple and intuitive approach is to employ time division multiple access (TDMA) technologies to allow multiple users to access the ES in turn. Furthermore, from the communication perspective, code division multiple access (CDMA) or non-orthogonal multiple access (NOMA) schemes are also possible candidates to support the multi-user MEG. However, it is worth noting that in the multi-user scenario, MEG not only needs to consider communication challenges, but also the model accessibility, resource allocation, and even users' personalized demands.

\section{Seed Transmission: Dynamic Power Allocation}\label{section:4}

An inspection of the seed design in Section III reveals that each seed ought to be regarded as a code block. It is essential that each seed is fully received and input into the compression decoder in its entirety. Nevertheless, in practical transmission, the requisite transmission time for a single code block may span multiple, or even dozens, of coherent fading channel blocks. Thus, the transmission of the generation seed in practical is likely to be given by
\begin{align}\label{y2}
 {{\bf y}_t} = h_t{{\bf x}_t} + {{\bf n}},
\end{align}
where ${\bf y} = [{\bf y}_0, {\bf y}_t,..., {\bf y}_T]$ and ${\bf x} = [{\bf x}_0, {\bf x}_t,..., {\bf x}_T]$. The utilization of channel equalization technology does not guarantee uniform interference levels for the content transmitted in each time slot, due to the inherent variability in channel fading strengths. Unfortunately, interference with a portion of the seed is likely to result in imperfect compression decoding and thereby affect the quality of the generated image as this factor is not considered in the training object $\mathbf{P1}$. Thus, in order to address the challenges posed by the evolving nature of fading channels, and optimize the overall quality of the generated images, we propose a power allocation scheme based on DRL.

\subsection{Problem Formulation for the Seed Transmission}

Further develop equation \eqref{y2}, the seed transmission model having dynamic power allocation is given by
\begin{align}\label{y3}
 {{\bf y}_t} = h_t p_t{{\bf x}_t} + {{\bf n}},
\end{align}
where $p_t$ represents the power allocated to the coherent fading time block $t$.
In order to minimize the impact of varying channels and noise during transmission on the final generated image, the DRL agent intelligently allocates the limited power budget for one generation task to each time slot. The FID score mentioned above will be applied as a standard to evaluate the quality of the generation. Then, the problem for the DRL agent assisted power allocation is given by
\begin{subequations}
\begin{align}
\mathbf{P2}: &\min_{p_t}   \text{FID}(\mathbf{I}^0, \mathbf{I}^g), \\
\textrm{s.t.} \ \
& \mathcal{T}_{\mathbf{I}^g} = \mathcal{T}_{\mathbf{I}^0}, \label{OPP2}\\
& \sum\limits_{t = 0}^{T} p_t \leq p_\text{max}, \label{OPP3}\\
& 0 < |\mathbf{x}_{\mathbf{I}^g}| < f_d^{-2}|\mathbf{x}_{\mathbf{I}^c}|, f_d > 1  ,
\end{align}
\end{subequations}
where constraint (\ref{OPP2}) is a constraint on the denoising process of the LDM sampler, which guarantees that the ground truth $\mathbf{I}^0$ and the generated image $\mathbf{I}^g$ has the same denoising steps, where $\mathcal{T}$ represents the maximum denoising steps. Constraint (\ref{OPP3}) ensures that the total power consumption for the transmission of image ${\mathbf{I}^g}$ is finite, where $T$ represents the transmission period.

\begin{figure}[t!]
\centering
\includegraphics[width=0.5\textwidth]{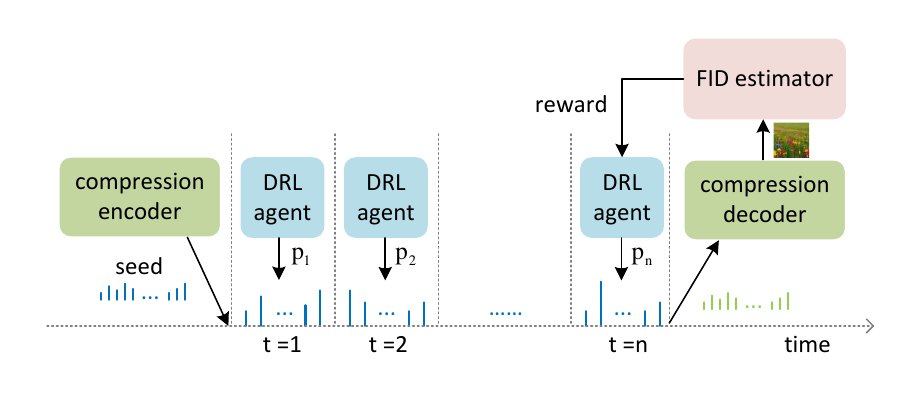}
\caption{Compression encoder for feature transmission.}
\label{Fig.6}
\end{figure}

\subsection{MDP of the MEG Transmission}

In the MEG EI protocol considered, the DRL agent responsible for power allocation is deployed on the ES side. As displayed in Fig. \ref{Fig.6}, the DRL agent's functionality is initiated upon the acquisition of the signal $\mathbf{x}$ that is to be transmitted.  Each fading block is regarded as a time slot for DRL, and the state observation of time slot $t$ can be denoted by $\bm{s}_t$. The DRL agent bears the responsibility of determining the action $a_t$ indicating transmission power for each fading block. Each transmission and generation process $0<t<T$ is treated as an episode. Once the generation of the image is complete, its FID score will be calculated and employed as the foundation for determining the reward $r_t$. Thus, the MDP $(\bm{s}_t, a_t, \bm{s}_{t+1}, r_t )$ is formulated. It is worth noting that since the power consumption in a period of time is a cumulative value, the MDP problem is a long-term optimization problem, so the DRL algorithm is employed to solve it. The design of state, action, and reward functions will be introduced below respectively.

\subsubsection{State Space}

The optimization of long-term power allocation strategies for multiple fading blocks, in a context of constrained power consumption, necessitates the integration of three distinct types of information. Given the transmission task of a specific image, the information that contributes to the decision for each time slot includes the channel state information $h_t$, remaining power budget $p_\text{max} - \sum\limits_{0}^{t} p_t$, and the piece of the coded feature $\mathbf{x}_t$ to be transmitted. This information needs to be included in the  state space, and it can be given by
\begin{align}\label{S}
\mathbf{s}_t = \{\mathbf{x}_t, {h}_{t}, p_\text{max} - \sum\limits_{0}^{t} p_t\}.
\end{align}

\subsubsection{Action Space}

Since the purpose of the DRL agent is to optimize power allocation, the action space only needs to contain the normalized power for the given time slot. The action is denoted by $a_t$. In order to precisely control the power, $a_t$ is set as a continuous variable and follows $a_t\in [0,1]$. For an output $a_t$, the power for time slot t is
\begin{align}\label{S}
p_t =  a_t \cdot p_\text{max}.
\end{align}

It is worth noting that in order to ensure the power constraint \eqref{OPP3}, when the allocated power of a time slot exceeds the remaining power, all the remaining power will be allocated to this time slot. Thus, the power allocation is given by
\begin{align}\label{S2}
p_t = p_\text{max} - \sum\limits_{0}^{t} p_t, \text{if  } a_t \cdot p_\text{max} > p_\text{max} - \sum\limits_{0}^{t} p_t.
\end{align}

\begin{algorithm}
\caption{DRL algorithm for the power allocation}
\label{LRISPPO}
\begin{algorithmic}[1]
        \STATE \textbf{Initialization:}Initialize the actor network $\varpi_{a,0}$, critic network $\varpi_{c,0}$.
        \FOR{each episode}
        \FOR{done = False}
        \STATE Choose $a_t$ according to the current state $\boldsymbol{s}_t$ and policy $\pi _{\varpi_\text{old} }\left ({a_{t}| \bm {s}_{t}}\right)$.
        \STATE Check the remaining power to determine the legality of the action $a_t$.
        \IF {Remaining power is sufficient}
            \STATE Execute action $a_t$.
        \ELSE
            \STATE Spend all remaining power.
        \ENDIF
        \IF {All symbols in $\mathbf{x}$ are transmitted}
            \STATE Decode image according to.
            \STATE Calculate reward $r_t$, done = True.
        \ELSE
            \STATE reward $r_t = 0$, done = False, and $\boldsymbol {s}_{t} \rightarrow \boldsymbol {s}_{t+1}$.
        \ENDIF
        \STATE Record MDP tuple $(\boldsymbol {s}_t,{a}_t,r_t,\boldsymbol {s}_{t+1})$.
        \ENDFOR
        \FOR{Each training epoch}
        \STATE Sample the MDP tuples and calculate loss according to \eqref{PPOL}.
        \STATE Train actor/critic networks according to \eqref{PPOback}.
        \ENDFOR
        \ENDFOR
\end{algorithmic}
\end{algorithm}

\begin{table*}[t!]

 \caption{Simulation Parameters}\label{SP}
 \centering
 \footnotesize
 \renewcommand\arraystretch{1.5}
 \begin{tabular}{|c|c|c|c|c|c|}
  \hline
  Parameter & Description & Value &   Parameter & Description & Value \\
  \hline
  $f_\text{c}$ & compression rate & 0.5 & $f_d$ & down sampling factor & 8  \\
  \hline
  $T$ & sample steps  & 50 & $H$ & image hight & 512 pixels \\
  \hline
  $W$ & image width & 512 pixels & $C$ & number of latent channels & 4 \\
  \hline
  lr & learning rate & 1e-3 &  ep & training epoch & 200 \\
  \hline
  $e$ & batch size & 1 samples & $\tau$ & training SNR & 20dB \\
  \hline
 \end{tabular}
\end{table*}

\begin{figure*}[t]
\centering
\includegraphics[width=0.8\textwidth]{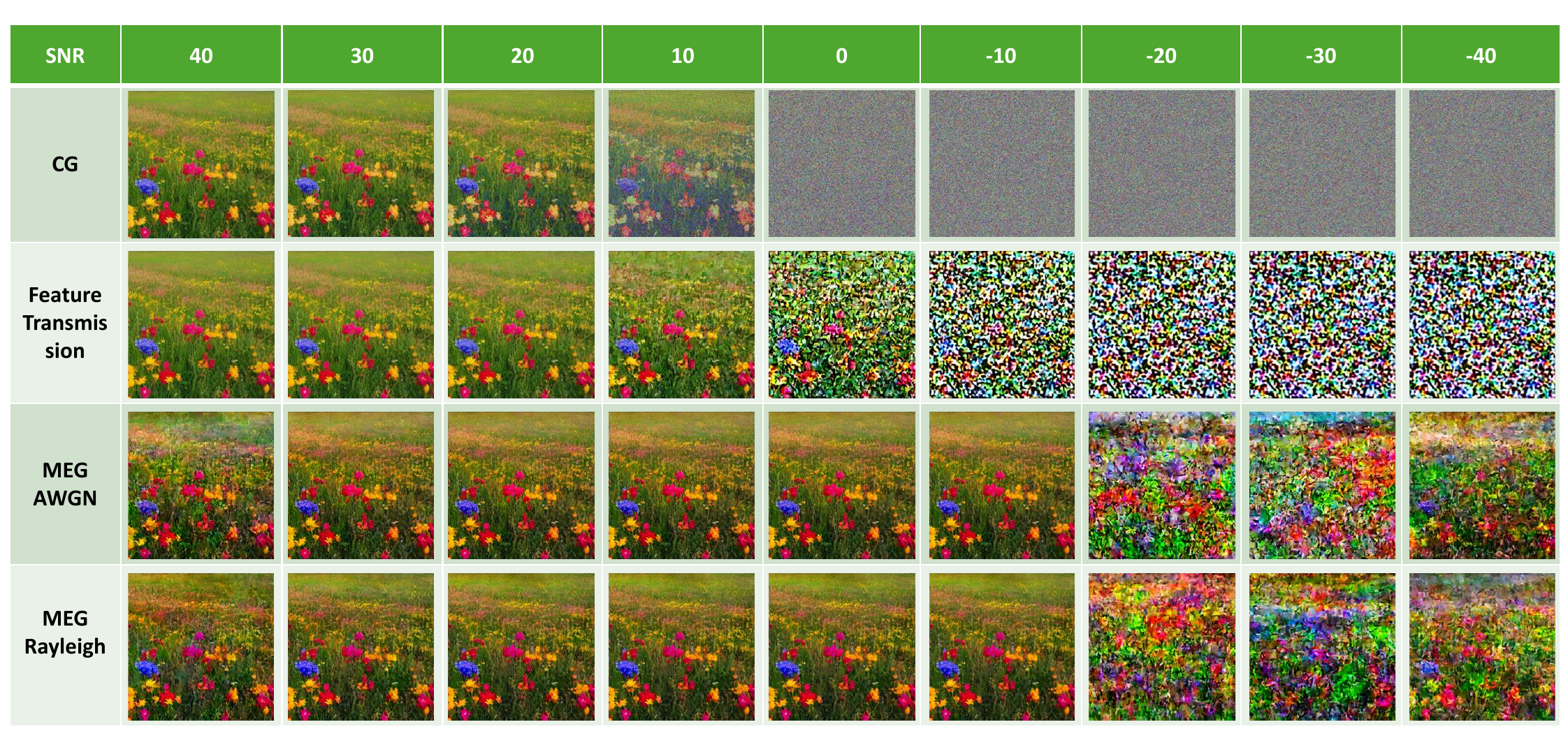}
\caption{Visualization examples for different generation schemes.}
\label{Fig.10}
\end{figure*}

\subsubsection{Reward Function}

In order to minimize the FID score of the decoded image. The reward is completely determined by the FID score obtained at the end of transmission. Therefore, the reward function of the proposed power allocation approach is given by

\begin{align}\label{S}
r_t =
\begin{cases}
0 , \text{if  } t < T, \\
-\text{FID}_{\mathbf{I}^g}  , \text{if  } t = T.
\end{cases}
\end{align}

\subsection{Agent Training}

In order to achieve an efficient optimization for power control, a policy optimization algorithm (PPO) is employed due to its outstanding performance among DRL algorithms. The aim of training the PPO agent is to obtain a stochastic policy $\pi _{\varpi }\left ({{a}_{t}| \bm{s}_{t}}\right)$, where $\varpi$ represents the parameter set of the neural network model. PPO agent employs the surrogate objective to update the policy, and it is given by
\begin{align}
L^\text{SO} = \mathbb {E} \left[{\displaystyle u_t(\varpi){A}_t}\right],
\end{align}
where $A$ represents the advantage function given in \cite{schulman2017proximal}, and
\begin{align}
u_{t} \left ({\varpi }\right)=&\frac {\pi _{\varpi }\left ({ {a}_{t}| \boldsymbol {s}_{t}}\right)}{\pi _{\varpi _{\mathrm{ old}}}\left ({ {a}_{t}| \boldsymbol {s}_{t}}\right)}.
\end{align}

In order to circumvent the extensive policy update that is likely to impede stable convergence, the loss function of the PPO algorithm is designed as
\begin{align}\label{Lt}
L_{t}\left ({{\varpi }}\right)=&\mathbb {E} \left [{L_{t}^{\mathrm{ CLIP}}\left ({{\varpi }}\right)-c_{1}L_{t}^{\mathrm{ VF}}\left ({{\varpi }}\right)}\right. \left.{\vphantom {\left [{L_{t}^{\mathrm{ CLIP}}\left ({{\varpi }}\right)-c_{1}L_{t}^{\mathrm{ VF}}\left ({{\varpi }}\right)}\right.}+\,\, c_{2}S\left [{\pi _{\theta }}\right]\left ({\boldsymbol {s}_{t}}\right)}\right],
\end{align}

The first term in \eqref{Lt} is the clipped surrogate objective, which is given by
\begin{align} \label{PPOL}
L_{t}^{\mathrm{ CLIP}}\left ({{\varpi }}\right)=&\mathbb {E} \left [{\min \left ({u_{t}^{s}\left ({\varpi }\right){A}_{t},{\textrm {clip}}\left ({u_{t}^{s}\left ({\varpi }\right),1-\epsilon, 1+\epsilon }\right){A}_{t}}\right)}\right].
\end{align}

The the probability ratio of the surrogate objective is clipped by the legal interval $[1-\epsilon, 1+\epsilon]$, where $\epsilon$ is a hyperparameter to adjust the range of the clip.

The second term $L_{t}^{\mathrm{ VF}}$ represents the squared value function error term, and $c_1$ is constants. $L_{t}^{\mathrm{ VF}}$ can be given by
\begin{align}
L_{t}^{\mathrm{ VF}}\left ({\varpi }\right)=&\mathbb {E}_{t}\left [{\left ({V_{\varpi }\left ({\boldsymbol {s}_{t}}\right)-V_{\pi _{\varpi _{\mathrm{ old}}}}\left ({\boldsymbol {s}_{t}}\right)}\right)^{2}}\right],
\end{align}

Finally, the third term in \eqref{Lt} represents the entropy bonus to accelerate the exploration.

The policy gradient $\nabla _{\omega }L_{t}\left ({\omega }\right)$ can be calculated by loss $L_{t}\left ({{\varpi }}\right)$, the parameter set can be updated following
\begin{equation}\label{PPOback}
 \varpi \leftarrow \varpi +\alpha\nabla _{\varpi }L_{t}\left ({\varpi }\right),
 \end{equation}
where $\alpha$ represents the learning rate.

\section{Numerical Results and Analysis}\label{section:5}

In the simulations, we tested the generated image quality and transmission overhead of the proposed MEG scheme. The LDM model developed by Stable Diffusion is employed as the basic image generation model, and the pre-trained checkpoint stable-diffusion-v1-5 is used for the parameters. Since this is an initial work of MEG, two benchmarks are invoked for the compression.  The first benchmark is the centralized generation, following the description in Section II. The other benchmark is the MEG transmitting raw latent features, where the compression encoder and decoder are not invoked in order to identify the gain of the compression coding. The detail of the parameter of the setup is given in Table \ref{SP}.

\begin{figure}[t]
\centering
\includegraphics[width=0.45\textwidth]{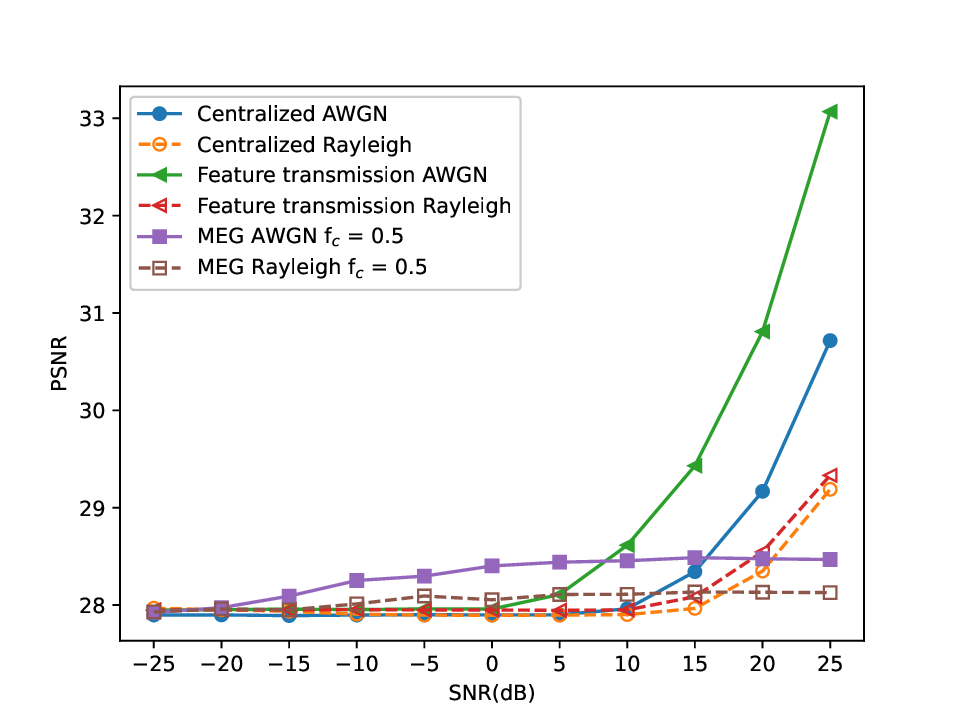}
\caption{PSNR performance of different generate approach (higher is better).}
\label{Fig.11}
\end{figure}

\begin{figure}[t]
\centering
%\vspace{-0.2cm}
\includegraphics[width=0.45\textwidth]{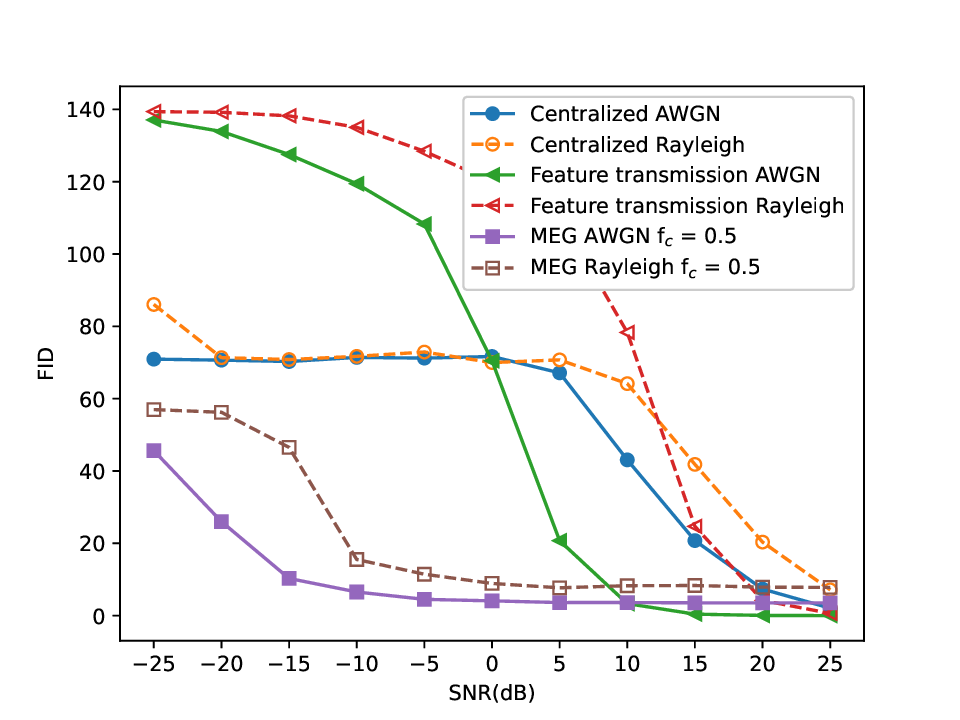}
\caption{FID performance of different generate approach (lower is better).}
\label{Fig.12}
\end{figure}

\begin{figure}[t]
\centering
\vspace{-0.2cm}
\includegraphics[width=0.5\textwidth]{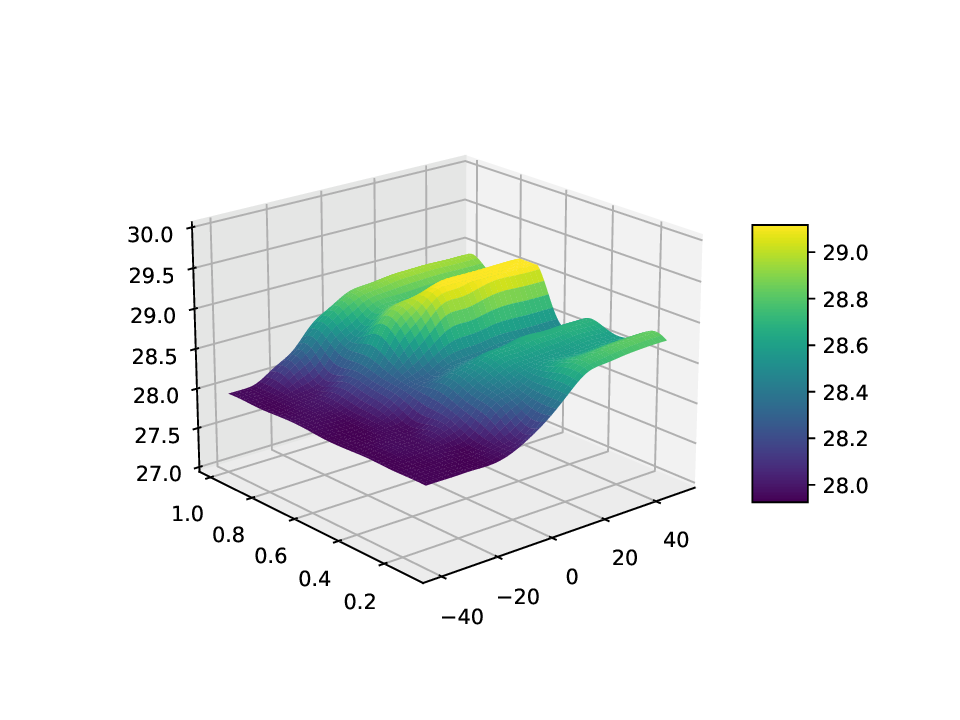}
\caption{PSNR score for different compression rate.}
\label{Fig.13}
\end{figure}

\begin{figure}[t]
\centering
\vspace{-0.2cm}
\includegraphics[width=0.5\textwidth]{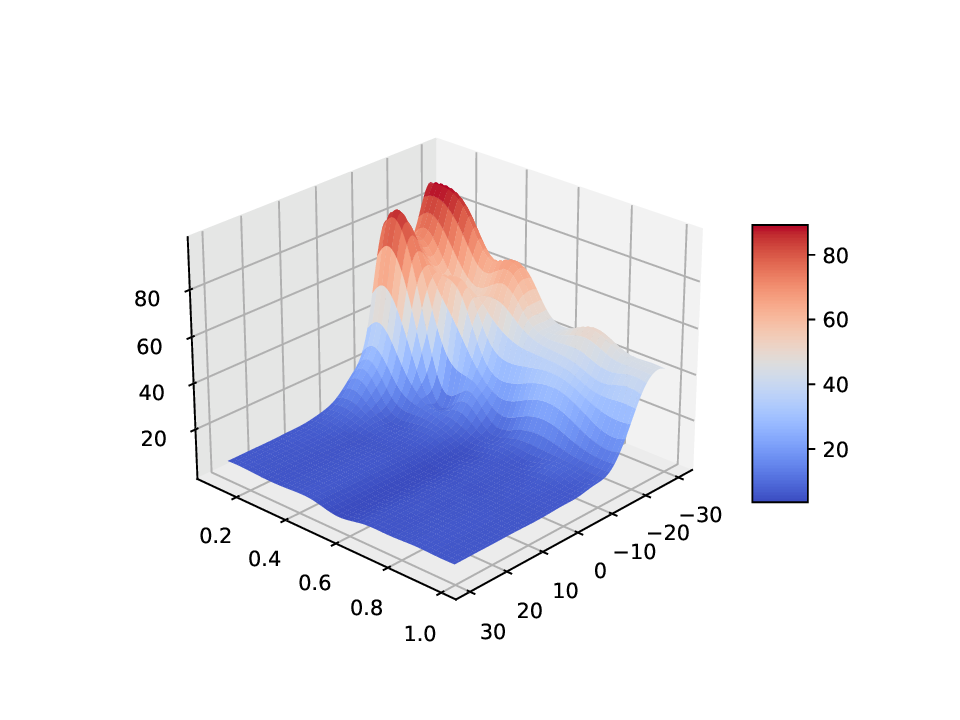}
\caption{FID score for different compression rate.}
\label{Fig.14}
\end{figure}

\subsection{Performance of Seed Enabled Image Generation}

Fig. \ref{Fig.10} presents a number of visible images generated over different schemes and SNR. The prompt of these images is 'colorful flowers on a field'. For the centralized generation scheme, the results exhibit classic pixel noise. There is a significant decrease in image quality when SNR=10db, and the image is almost completely invisible when SNR=0db. Direct transmission of unencoded latent features is also a possible solution. In this case, although the amount of data transmitted is greatly reduced, the quality of the generated image is not inferior to directly transmitting the generated image. Observing the image generated when SNR=10db, the transmission error caused the background to be blurred, but the image of the flowers in the front row was basically restored correctly. Under the condition of SNR=0db, some flowers in the front are vaguely visible, and this performance is also better than the center generation solution. For the MEG scheme, we show the renderings of two encoding compression ratios: $f_c=0.5$ and $f_c=0.9$. It can be found that in general, the quality of the generated image when $f_c=0.9$ is better than $f_c=0.5$. Under low SNR conditions, MEG demonstrates competitive generation capabilities. At SNR below -20dB, a large number of errors can be observed, but these errors appear as semantic errors rather than pixel errors. Specifically, MEG eventually generated some chaotic flowers under low signal-to-noise ratio conditions, or even a superimposed state of multiple images, but the pixel errors were greatly alleviated.

The PSNR of the generated image from different generation schemes is compared in Fig. \ref{Fig.11}. For both AWGN and Raleigh fading channel, under low SNR conditions, the MEG scheme has a small advantage. However, after the SNR is larger than 10db, the PSNR of directly transmitted images or latent features rises rapidly beyond MEG. The disadvantage of the MEG scheme at high SNR comes from the DNN-based encoder and decoder. Due to the existence of over-fitting, DNN-based encoders and decoders cannot be trained infinitely. Therefore, even under noise-free conditions, the loss of DNN encoding and decoding cannot approach 0 due to the limited training process. It can be inferred that DNN-based compression coding itself introduces noise. However, this problem can be avoided by using other encoding and decoding methods. Nevertheless, the benefits of DNN encoding and decoding in low SNR scenarios will be lost.

\begin{table*}[t!]
\caption{Communication Overhead and Model Complexities for MEG}\label{table}
\centering
\footnotesize
\renewcommand\arraystretch{1.5}
\begin{tabular}{|c|c|c|c|c|c|c|c|}
\hline
Generation mode                & \begin{tabular}[c]{@{}c@{}}Centralized \\ generation\end{tabular} & \begin{tabular}[c]{@{}c@{}}Feature \\ transmission\end{tabular} & \begin{tabular}[c]{@{}c@{}}MEG \\$f_c=0.1$ \end{tabular} & \begin{tabular}[c]{@{}c@{}}MEG \\$f_c=0.3$ \end{tabular} & \begin{tabular}[c]{@{}c@{}}MEG \\$f_c=0.5$ \end{tabular} & \begin{tabular}[c]{@{}c@{}}MEG \\$f_c=0.7$ \end{tabular} & \begin{tabular}[c]{@{}c@{}}MEG \\$f_c=0.9$ \end{tabular} \\ \hline
Transmitted data & 1,048,576   & 16384                    & 1638       & 4915       & 8192       & 11469      & 14746      \\ \hline
Total model size & 7.17GB      & 7.17GB                   & 7.4GB      & 8.16GB     & 8.78GB     & 9.64GB     & 10.08GB    \\ \hline
\end{tabular}
\end{table*}

Fig. \ref{Fig.12} presents the FID performance for the different generation schemes. Overall, the FID performance of the proposed MEG demonstrates the same general trend as PSNR. The MEG is unlikely to perfectly recover the same image of the perfect image under SNR conditions, but it obtained a significant advantage under low SNR conditions compared with conventional generation schemes. Specifically, the FID score of the MEG scheme only shows a significant increase when SNR is under -20dB. The simpler scheme of directly transmitting features is also better than the centralized generation scheme in terms of FID score. The FID loss of directly transmitting features increases significantly at 0dB-10dB, while the centralized generation scheme starts FID attenuation at 20dB, which is consistent with the visualization results in Fig. \ref{Fig.10}.

The PSNR performance of MEG versus both compression rate $f_c$ and SNR is displayed in Fig. \ref{Fig.13}, where the x and y axis represent the compression rate and the SNR, respectively. As a general rule, the PSNR decreases as the compression ratio decreases. However, different from the conventional coding technologies, the most significant impact on the performance is not only the code length but also the parameter sets $\bm{\omega}$ and $\bm{\psi}$. Taking the compression rate of 0.8 as an example, the PSNR performance at this time is stronger than the compression rate of 0.7 or 0.9 forming a PSNR peak. The reason is that we conducted additional experiments and more accurately tuned the hyperparameters of the compression decoder for the compression ratio of 0.8.

Fig. \ref{Fig.14} displays the FID score against the compression rate $f_c$ and SNR. Compared with PSNR, the MEG model has greater discrimination in FID scores. FID scores are remarkably high under low SNR and low compression rate conditions. This result is consistent with the general transmission interference and coding theory. On the one hand, tn low SNR conditions, the decreasing trend of FID score with the increase of code length is more significant than the trend of PSNR. It can be observed that the FID score is generally high in the interval of $f_c<0.4$, staying medium with slight fluctuation in the interval of $<0.4f_c<0.8$. On the other hand, the FID score is sensitive to the SNR, and a clear threshold at -20dB can be observed. When the signal-to-noise ratio is greater than -10dB, the FID score can be basically guaranteed to be at a low level, which means that the proposed scheme can work under common communication conditions.

\subsection{Complexity and Communication Overhead of MEG}

In addition to the performance of generated images, communication overhead and model size are also two key indicators as well. The model size is directly related to whether it is feasible to deploy MEG models on edge devices. Table \ref{table} indicates the total model size and the amount of float numbers required to be transferred under different generation approaches. Observing the transmission data amount, the first conclusion is that using MEG, even directly transmitting unencoded latent features can significantly reduce the amount of data that needs to be transmitted, which is in line with the description in \textbf{Remark 1}. In terms of model size, using the central generation scheme only requires the GAI model, and it has minimal model complexity. The MEG schemes have different increases in model size due to differences in the structure of the compression encoder and decoder employed for different compression rates. Due to the increase in encoding length, increasing network size leads to increases in encoding and decoding complexity. However, this complexity increase is acceptable overall as these sub 10GB models are feasible to deploy at ES and UE.

\subsection{Performance of the DRL-enabled Power Allocation}
\begin{figure}[t]
\centering
\vspace{-0.2cm}
\includegraphics[width=0.5\textwidth]{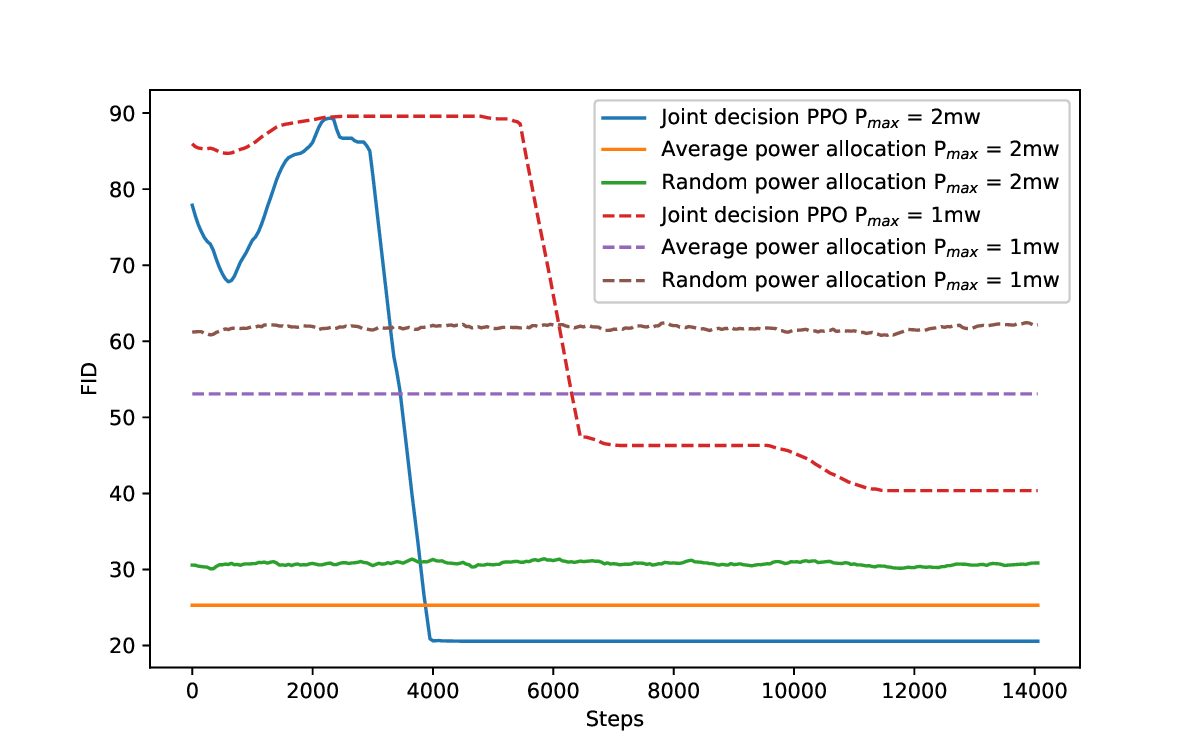}
\caption{FID score against different power constraint.}
\label{Fig.15}
\end{figure}
The FID scores for the various power allocation schemes are presented in Fig. \ref{Fig.15}. Initially, in the context of evolving fading channels, the DRL algorithm demonstrates robust convergence following training. Secondly, in comparison with the scheme utilizing solely tie power allocation, the employment of the DRL algorithm has been observed to result in a notable reduction in FID score, particularly under disparate power constraints. In the case of 1 mW, where power is more limited, the FID gain from DRL enabled power allocation reaches 20.

\vspace{-0.2cm}
\section{Conclusions}\label{section:6}

An MEG model was proposed in this paper, which enables users to obtain generated content from the edge devices in the network.  MEG deploys the GAI model in a distributed manner on ES and UE to jointly complete the generation task, thereby reducing data transmission consumption and improving noise resistance. The proposed MEG scheme was discussed in two parts, namely seed production and seed transmission, respectively. To product the generation seed, we developed a MEG model for text-to-image generation based on the LDM model and compression coding. A deep learning-based compression encoder and decoder were proposed, and the coded features were transmitted to UE instead of images. To against the fading channel, a DRL algorithm was invoked to control the transmission power allocation over fading blocks to optimize the quality of the image. The simulation results show that 1) the proposed MEG scheme has the capability to improve image quality under low SNR conditions; 2) the DRL approach can reduce the FID score for the generated image by intelligently allocating the transmit power for the generation seed; 3) MEG schemes can exponentially reduce the communication resources required for GAI.

%Since the development of MEG is still in its initial stage, there are still a number of open research problems that need to be solved, including model deployment, multi-ES parallel processing, latency performance optimization, and etc.

\section*{Appendix A \\Specification of the Neural Network and Parameters Number }

\begin{table}[htb]
\caption{Structure of the example compression encoder and decoder for $f_c=0.5$.}
\centering
\footnotesize
\begin{tabular}{|p{.14\textwidth}|p{.19\textwidth}|p{.07\textwidth}|}
\hline
{ Model}                                 & { Layer}                                          & { Param}       \\ \hline
{ }                                      & { Flatten}                                        & { /}           \\ \cline{2-3}
\multirow{-2}{*}{{ Compression encoder}} & { Linear(in\_features=16384, out\_features=8192)} & { 134,225,920} \\ \hline
{ }                                      & { Relu(in\_features=8192, out\_features=16384}    & { 134,234,112} \\ \cline{2-3}
{ }                                      & { Normalize(in\_features=16384)}                  & { /}           \\ \cline{2-3}
{ }                                      & { Relu(in\_features=16384, out\_features=9000}    & { 147,465,000} \\ \cline{2-3}
{ }                                      & { Normalize(in\_features=9000)}                   & { /}           \\ \cline{2-3}
{ }                                      & { Relu(in\_features=9000, out\_features=16384}    & { 147,472,384} \\ \cline{2-3}
\multirow{-6}{*}{{ Compression decoder}}  & {LayerNorm(16384, eps=1e-06)}                    & { 32,768}      \\ \hline
{ Total}                                 & {}                                               & { 563,430,184} \\ \hline
\end{tabular}
\end{table}

\renewcommand{\baselinestretch}{1.0}
\bibliography{ref_abb}
\bibliographystyle{IEEEtran}
\end{document}